\RequirePackage{fix-cm}
\documentclass[smallextended]{svjour3}       
\smartqed  
\usepackage{graphicx}
\def\beq{\begin{equation}}
\def\eeq{\end{equation}}
\def\bea{\begin{eqnarray}}
\def\eea{\end{eqnarray}}
\def\a{\alpha}
\def\b{\beta}

\def\m{\mu}
\def\n{\nu}
\def\h{\eta}

\def\G{\Gamma}
\def\F{\Phi}
\def\hh{\tilde{h}}

\def\da{\partial^2}
\def\aa{\partial}
\newcommand{\an[1]}{\partial_#1}
\def\2{\;\;}
\def\4{\;\;\;\;}

\begin{document}

\title{From GEM to Electromagnetism}


\author{A. Bakopoulos  \and P. Kanti }


\institute{A. Bakopoulos \at
              Department of Physics,
University of Ioannina, Ioannina GR-45110, Greece \\
              \email{bakop2@hotmail.com}           
           \and
           P. Kanti \at
              Division of Theoretical Physics, Department of Physics,
University of Ioannina, Ioannina GR-45110, Greece\\
              Tel.: +30-26510-08486\\
              Fax: +30-26510-08698\\
              \email{pkanti@cc.uoi.gr}
}

\date{Received: date / Accepted: date}

\maketitle

\begin{abstract}
In the first part of the present work, we focus on the theory of 
gravitoelectromagnetism (GEM), and we derive the full set of equations
and constraints that the GEM scalar and vector potentials ought to satisfy.
We discuss important aspects of the theory, such as the presence
of additional constraints resulting from the field equations and gauge
condition, the requirement of the time-independence of the vector potential
and the emergence of additional terms in the expression of the Lorentz force.
We also propose an alternative ansatz for the metric perturbations that is
found to be compatible only with a vacuum configuration but evades several
of the aforementioned obstacles. In the second part of this work, we pose
the question of whether a tensorial theory using the formalism of General
Relativity could re-produce the theory of Electromagnetism. We demonstrate
that the full set of Maxwell's equations can be exactly re-produced for a
large class of models, but the framework has several weak points common
with those found in GEM.
\end{abstract}


\section{Introduction}
\label{Intro}
One of the most ambitious objectives of Theoretical Physics is the 
formulation of a theory that would unify all forces in nature. 
Maxwell's theory of electromagnetism paved the way by unifying two,
apparently distinct, sets of phenomena in nature associated with the
electric and magnetic field, respectively. Decades later another
unification was achieved, this time of electromagnetic and weak interactions
in the context of Standard Model of particle physics.
Although the true unification of forces met at a microscopic level,
i.e. electromagnetic, weak and strong, will only take place -- if it
happens at all -- at a much higher energy scale, the use of a common
formalism, based on gauge theories, has already imposed a notion
of `unification' in people's minds and certainly has deepened our
understanding of particle interactions. 

Gravity, on the other hand, is still resisting our attempts to
implement it in a common framework with all other forces. Ironically,
it was the profound similarity of the Newtonian and Coulomb potential,
noticed centuries ago, that generated the idea of unification of forces
at the first place. Even at the level of fields, the corresponding 
equations, namely\footnote{Throughout this work, we follow the
conventions of \cite{Landau}.}
\begin{equation}\label{1}
\vec{\nabla}\cdot\vec{E} = 4\pi\rho_E\,, \qquad \qquad
\vec{\nabla}\cdot \vec{g}=-4\pi G\rho_M\,,
\end{equation}
where ($\vec{E}$, $\vec{g}$) and ($\rho_E$, $\rho_M$) are the electric
and gravitational fields and their corresponding charge and mass densities,
clearly show the similarity in how electrostatics and gravity work.
The counter-argument that the similarity breaks down as soon as we
introduce the magnetic field was invalidated by the discovery of the
Lense-Thirring effect \cite{LT} where the gravitational field of a
rotating body may be expressed in terms of an additional `vector
potential' which at large distances creates a gravitational `magnetic'
dipole due to the mass current. 

The first attempt to unify gravity with electromagnetism, the Kaluza-Klein
theory, used the additional degrees of freedom of a higher-dimensional
gravitational theory to accommodate the necessary gauge degrees of freedom.
The theory was purely geometrical and revealed the relation between 
gauge invariance and coordinate transformation. The same idea was used
much later in the formulation of string and M-theory \cite{Witten}\cite{Duff}.
Together with loop quantum gravity \cite{Ashtekar}, these theories
are the primary candidates for the unification of gravity with the other
forces. However, a complete unification has not yet been achieved, and, as a
result, several other attempts to derive a theory of electromagnetism, either
classical or quantum, from a geometric theory have appeared in the literature.
Among others, these include attempts to derive electromagnetism from General
Relativity itself \cite{Evans}\cite{Dadhich}\cite{Agut}, 
from variants of General Relativity \cite{Anastasovski},
from the Born-Infeld-Einstein theory \cite{Vollick}
or from other geometric theories
\cite{Weingarten}\cite{Kassandrov}\cite{Unzicker}\cite{Gonzalez}\cite{Torrome}.

Motivated by the clear analogy between gravity and electromagnetism,
supported further by the results of the work by Lense and Thirring,
another idea, that of gravitoelectromagnetism (GEM) \cite{GEM}\cite{Mashhoon},
has developed over the years. The idea amounts to studying gravity in the
context of General Relativity by using the terminology of electromagnetism.
For example, for a rotating gravitational source, the metric describing
the spacetime around it may be expressed in terms of a scalar quantity
$\Phi$ and a vector $\vec{A}$, that at large distances are associated
to the Newtonian potential and angular momentum of the source. This
idea has helped to investigate further the analogy between the two
theories but also to understand gravity better by looking for 
gravitational analogues of electromagnetic phenomena in the context
of General Relativity. 


The first part of the present work focuses on the theory of gravitoelectromagnetism.
Following the standard assumptions and conventions, we express the metric
perturbations in terms of a scalar and vector potential, and we 
derive the full set of equations and constraints that these two quantities
ought to satisfy. Apart from the set of equations that exhibit a close
similarity to Maxwell's equations, we derive a number of additional constraints
from the remaining components of Einstein's field equations. In addition, we
demonstrate that the analogy between gravity and electromagnetism holds only under
the assumption that the vector potential is time-independent; this feature
is dictated also by the additional components of the transverse gauge
condition, that is implemented in the standard form of the theory.
We also re-derive the expression of the
analogue of the Lorentz force as this follows from the geodesic equation:
although we fully agree on the form of the terms appearing in previous
works on GEM, we show that extra corrections emerge not all of which can
be ignored in the classical limit. 

In an effort to explore different directions in the context of gravitoelectromagnetism,
we then deviate from the standard assumptions and we propose an alternative ansatz
for the metric perturbations. The novel feature of this ansatz is that all
components of the metric perturbations are assumed to be non-zero. In the
context of GEM, where a specific form of the energy-momentum tensor is assumed,
the spatial components of the metric perturbations are indeed significantly
suppressed and thus ignored. Our ansatz, on the other hand, turns out to be
compatible with a vacuum configuration where indeed all components can be
of the same magnitude. We show that this ansatz is accompanied by a number
of attractive features: any additional constraints from the field equations 
trivially vanish while the geodesics equation takes the exact form of the
equation of the Lorentz force with no extra corrections. The use of
alternative ansatze for the metric perturbations, such as the one studied
here, could perhaps open new directions of thinking and lead to a novel
class of phenomena in the context of gravitoelectromagnetism. 

The similarity that the perturbed Einstein's field equations have with those
of electromagnetism naturally leads to the question of whether one could
re-produce the true electromagnetism from a tensorial theory with a formalism
similar to that of the General Theory of Relativity. In the second part of
our paper, we attempt to answer this question. The starting point of our
analysis are again Einstein's field equations satisfied by the metric
perturbations at a linear approximation -- in the context of our analysis,
however, the field equations will be modified to allow for the substitution
of the gravitational constant by an, initially, undefined one; demanding 
that particular components of the field equations reduce to Maxwell's
equations, this constant will be determined solely in terms of the velocity
of light. A scalar $\hat \Phi$ and a vector potential $\hat{\vec{A}}$ are also
introduced through the components of the metric perturbations. However,
contrary to the case of gravitoelectromagnetism, these potentials will be
assumed to be the true potentials of electromagnetism, and not `components'
of the gravitational field. We employ a general ansatz for the form of
the metric perturbations, and we demonstrate that indeed the full set of
exact Maxwell's equations can be reproduced for a large class of choices.
However, the ensuing model of electromagnetism suffers from a number of 
weak points which we discuss in detail. 

Let us clarify that our analysis in the second part of our paper does
not propose a unification theory since gravity would in fact be absent.
Nor does it aim to replace the current theory of electromagnetism - afterall,
its quantum version, the Quantum Electrodynamics, is one of the most
successful and accurate theories in Physics. It rather aims at investigating
how far the analogy between gravity and electromagnetism, that was noted in,
and employed by, GEM, extends, and where and why it breaks down. We believe
that such an analogy, if found to work satisfactorily, would link the two
theories at a deeper level and assist, in a complimentary way, in the
formulation of a unified theory that is still missing. 

The outline of our paper is as follows: in section 2, we present the
theoretical framework and basic tools for our analysis. In Section 3, we 
focus on the theory of gravitoelectromagnetism: we perform a comprehensive
study of the full set of equations and constraints arising, and we re-derive
the equivalent of the equation for the Lorentz force; at the end, we propose
a novel ansatz for the metric perturbations and study its consequences. 
In Section 4, we turn to a tensorial theory that uses the formalism of
General Relativity and we attempt to re-produce the full set of Maxwell's
equations; we discuss in detail the successes and weak points of such an
approach. We present our conclusions in Section 5. 

%

\section{The Theoretical Framework}

Our starting point will be Einstein's field equations, and more
specifically the equations obeyed by the metric perturbation $h_{\mu\nu}$
defined through the relation 
\begin{equation}
g_{\mu\nu} (x^\mu) =\eta_{\mu\nu} + h_{\mu\nu}(x^\mu)\,,
\label{metric}
\end{equation}
where $\eta_{\mu\nu}$ is the Minkowski metric of the flat
spacetime\footnote{Throughout this work, we will use the $(+1,-1,-1,-1)$ signature 
for the Minkowski tensor $\eta_{\mu\nu}$.}  
and $x^\mu=(ct, \vec{x})$. In the context of General Relativity, the perturbations
$h_{\mu\nu}$ are assumed to be sourced by gravitating bodies and to obey the
inequality $|h_{\mu\nu}|\ll 1$ so that a linear-approximation
analysis may be followed. Here, we will work along the same lines - to this
end, we briefly review the corresponding formalism that leads to the equations
in the linear-order approximation (for a more detailed analysis, see for
example \cite{Landau}). Employing Eq. (\ref{metric}) and keeping only terms 
linear in the perturbation $h_{\mu\nu}$, we easily find that the Christoffel
symbols assume the form
\begin{equation}\label{Christoffel}
\Gamma^{\alpha}_{\mu\nu}=\frac{1}{2}\,\eta^{\alpha\rho}\left(
h_{\mu\rho,\nu}+h_{\nu\rho,\mu}-h_{\mu\nu,\rho}\right).
\end{equation}
The above leads, in turn, to the following expression for the Ricci tensor
\begin{equation}\label{RicciT}
R_{\mu\nu}=\frac{1}{2}\left({h^\rho}_{\mu,\nu\rho}+
{h^\rho}_{\nu,\mu\rho}-\da h_{\mu\nu} - h_{,\mu\nu}\right),
\end{equation}
where $\da=\h^{\mu\nu}\aa_\m\aa_\n$, and indices are raised and lowered by
the Minkowski metric $\eta_{\mu\nu}$. In the above, we have also defined
the trace of the perturbation $h=\h^{\m\n}h_{\m\n}$. Similarly, the Ricci
scalar is found to be
\begin{equation}\label{RicciS}
R={h^{\m\n}}_{,\m\n} - \da h\,.
\end{equation}
Combining all the above, the Einstein tensor, $G_{\m\n} \equiv R_{\m\n}-
\frac{1}{2}g_{\m\n}R$, takes the form
\begin{equation}\label{Einstein}
G_{\m\n}=\frac{1}{2}\left({h^\a}_{\m,\n\a}+ {h^\a}_{\n,\m\a}-
\da h_{\m\n}-h_{,\m\n}-\h_{\m\n}\,{h^{\a\b}}_{,\a\b}+\h_{\m\n}\,\da h \right).
\end{equation}
For simplicity, as is usually the case in gravity, we will also define the new
perturbations
\begin{equation}\label{newh}
\hh_{\m\n}=h_{\m\n} - \frac{1}{2}\,\h_{\m\n}\,h\,,
\end{equation}
in terms of which the Einstein tensor takes the simpler form
\begin{equation}\label{Einstein_new}
G_{\m\n}=\frac{1}{2}\left({\hh^{\a}}_{\2\m,\n\a}+{\hh^{\a}}_{\2\n,\m\a}-
\da \hh_{\m\n}-\h_{\m\n}\,{\hh^{\a\b}}_{\4,\a\b}\right).
\end{equation}
We will also assume that the above tensor satisfies Einstein's field equations,
i.e. $G_{\mu\nu}= k\,T_{\mu\nu}$, where $T_{\mu\nu}$ is the
energy-momentum tensor and $k$ a constant whose value will differ in the
first and second part of this work. Overall, our basic equations will
be the following ones
\begin{equation}\label{field_eqs}
{\hh^{\a}}_{\2\m,\n\a}+{\hh^{\a}}_{\2\n,\m\a}-
\da \hh_{\m\n}-\h_{\m\n}\,{\hh^{\a\b}}_{\4,\a\b}=2k\,T_{\mu\nu}\,.
\end{equation}
In what follows, we will assume that the distribution of energy in the
system is described by the expression $T_{\mu\nu}=\rho\,u_\mu u_\nu$,
where $\rho$ the charge density and $u^\mu=(u^0,u^i)=(c,\vec{u})$ the
velocity of the source.


\section{GravitoElectroMagnetism}

\subsection{The traditional ansatz for the perturbations}

In the context of the theory of gravitoelectromagnetism, Eqs. (\ref{field_eqs})
are Einstein's linearised gravitational field equations with $k=8\pi G/c^4$.
The components of the gravitational perturbations $\tilde h_{\mu\nu}$ have
the form \cite{Mashhoon}
\beq 
\tilde h_{00}=\frac{4\Phi}{c^2}\,, \qquad \tilde h_{0i}=-\frac{2A^i}{c^2}\,,
\qquad \tilde h_{ij}=0\,, \label{case1}
\eeq
and are expressed in terms of a scalar $\Phi(x^\mu)$ and a vector potential
$\vec{A}(x^\mu)$, the so-called GEM potentials. The $\tilde h_{00}$ component yields
the Newtonian potential $\Phi$, while the $\tilde h_{0i}$ component is associated
to the `vector' potential $\vec{A}$ generated by a rotating body; the $\tilde h_{ij}$
component is usually assumed to be negligible due to the suppression of the
corresponding source by a $1/c^4$ factor. To see the above, we may contract
Eq. (\ref{newh}) by $\eta^{\mu\nu}$ and find $h=-\tilde{h}$; this allows us to
write the inverse relation between the original and the new perturbations as 
\beq
h_{\mu\nu}= \tilde h_{\mu\nu} -\frac{1}{2}\,\eta_{\mu\nu}\,\tilde{h}\,.
\label{inversehnew}
\eeq
Then, by use of the definition (\ref{metric}), the spacetime line-element
assumes the form 
\beq
ds^2=c^2\left(1+\frac{2\Phi}{c^2}\right) dt^2-\frac{4}{c}\,(\vec{A} \cdot d\vec{x})\,dt
- \left(1-\frac{2\Phi}{c^2}\right) \delta_{ij} dx^i dx^j\,.
\label{line-element-GEM}
\eeq

For the evaluation of the field equations (\ref{field_eqs}),  we will need
the components of $\tilde h_{\mu\nu}$ in mixed form - these are:
\beq 
\tilde h^{0}_{\ 0}=\frac{4\Phi}{c^2}\,, \qquad \tilde h^{0}_{\ i}=-\frac{2A^i}{c^2}\,,
\qquad \tilde h^{i}_{\ 0}=\frac{2A^i}{c^2}\,,\qquad \tilde h^{i}_{\ j}=0\,.
\eeq
Then, using the above, the field equations (\ref{field_eqs})
reduce to the following system of equations
\bea
\frac{\delta^{ij}}{c^2}\,\partial_i \partial_j \Phi &=& \frac{k}{2}\,\rho\,u_0\,u_0 
\label{Poisson0GEM}\\[1mm]
\frac{1}{c^2}\,\partial_i\left(\frac{1}{2}\,\partial_k A^k +\frac{1}{c}\,\partial_t \Phi\right)
-\frac{1}{2c^2}\,\delta^{kl}\,\partial_k \partial_l A^i& =& \frac{k}{2}\,\rho\,u_0\,u_i 
\label{4Maxwell0GEM}\\[1mm]
-\frac{1}{2c^3}\,\partial_t \left(\partial_i A^j +\partial_j A^i\right)
+ \delta_{ij}\left[\frac{1}{c^4}\,\partial_t^2 \Phi + 
\frac{1}{c^3}\,\partial_t (\partial_k A^k)\right] &=& \frac{k}{2}\,\rho\,u_i\,u_j\,,
\label{ExtraGEM}
\eea
for $(\mu,\nu)=(0,0), (0,i)$ and $(i,j)$, respectively.

Adopting a more familiar notation and using $k=8\pi G/c^4$, Eq. (\ref{Poisson0GEM})
readily takes the analogue of Poisson's law
\beq
\nabla^2\,\Phi=4 \pi G \rho\,,
\label{PoissonGEM}
\eeq
while Eq. (\ref{4Maxwell0GEM}) in turn can be rewritten as
\beq
\vec{\nabla}\,\left[\vec{\nabla} \cdot \left(\frac{\vec{A}}{2}\right) + 
\frac{1}{c}\,\partial_t \Phi\right]
-\nabla^2 \left(\frac{\vec{A}}{2}\right) = \frac{4\pi G}{c}\,\rho \,\vec{u}\,.
\label{4MaxwellGEM}
\eeq
Defining, in analogy with electromagnetism, the GEM fields $\vec{E}$ and $\vec{B}$
in terms of the GEM potentials \cite{Mashhoon}
\beq
\vec{E} \equiv -\frac{1}{c}\,\partial_t \left(\frac{\vec{A}}{2}\right) -\vec{\nabla} \Phi\,, 
\qquad \vec{B} \equiv \vec{\nabla} \times \left(\frac{\vec{A}}{2}\right)\,, \label{EB_GEM}
\eeq
one may easily see that Eqs. (\ref{PoissonGEM}) and (\ref{4MaxwellGEM}) reduce to
\beq
\vec{\nabla} \cdot \vec{E} = 4\pi G \rho\,, \qquad 
\vec{\nabla} \times \vec{B}=\frac{1}{c}\,\partial_t \vec{E} + 
\frac{4\pi G}{c}\,\vec{j}\,, \label{FinalEqsGEM}
\eeq
respectively, where we have used the definition of the current vector $\vec{j}=\rho \vec{u}$.

The aforementioned discussion on the equations satisfied by the GEM fields $\vec{E}$
and $\vec{B}$, as these follow from the linearised Einstein's equations, first
appeared in \cite{Mashhoon}. However, the explicit form of the equations
(\ref{Poisson0GEM}-\ref{ExtraGEM}) obeyed by the GEM potentials $\Phi$ and 
$\vec{A}$ was not given. A careful inspection of 
Eqs. (\ref{Poisson0GEM}-\ref{4Maxwell0GEM}), or equivalently
(\ref{PoissonGEM}-\ref{4MaxwellGEM}),
reveals that these reduce indeed to the form of (\ref{FinalEqsGEM}) if and
only if the vector potential is static. Therefore, the constraint 
$\partial_t \vec{A}=0$, that in the context of the analysis of \cite{Mashhoon}
is often presented as merely a simplifying assumption, is in fact a direct
result of the analysis. 

The same result follows from the application of the gauge condition -- that
aims to remove some of the arbitrariness in the theory caused by its invariance
under coordinate transformations -- to the system of equations. The most
usual gauge condition, and the one used in the context of \cite{Mashhoon},
is the so-called {\it transverse} condition ${\hh^{\m\n}}_{\4,\n}=0$.
The time-component of the gauge condition takes the Lorentz form
\begin{equation}\label{Lor_GEM}
\frac{1}{c}\,\an[t]\F + \vec{\nabla}\cdot \left(\frac{\vec{A}}{2}\right)=0\,,
\end{equation}
in clear analogy to electromagnetism. Nevertheless, due to its tensorial
structure, the gauge condition has three more components that may be
collectively written as
\begin{equation}\label{gauge_add-GEM}
\frac{1}{c}\,\partial_t \vec{A}=0\,.
\end{equation}
The above constraint was not discussed in \cite{Mashhoon}, however, it is an
indispensable part following from the applied gauge condition that re-affirms
the necessary time-independence of the vector potential $\vec{A}$ (for different
approaches regarding the role of this constraint in the context of
gravito-electromagnetism, see 
\cite{Harris}\cite{Braginsky}\cite{Pascual}\cite{Costa}\cite{Natario}).

Let us now return to the linearised field equations (\ref{field_eqs}): their
spatial components lead to a third set of constraints, Eq. (\ref{ExtraGEM}),
that is largely ignored in the literature \footnote{While this manuscript was
at the stage of production, we were notified by the authors of 
\cite{Costa}\cite{Natario} that a study of these additional
equations was previously performed in their works.}. Its diagonal components
(i.e. for $i=j$) reduce to the relation
\beq
\partial_t^2 \Phi =- \frac{\pi}{3} \rho\,|\vec{u}|^2\,,
\label{diagonal-GEM}
\eeq
with $(u^1)^2=(u^2)^2=(u^3)^2$, while the off-diagonal ones (for $i \neq j$) give
\beq
\partial_0 \left(\partial_i A^j +\partial_j A^i\right)= 8\pi G\rho 
\,\frac{u_i u_j}{c^2}\,. \label{off-diagonal-GEM}
\eeq
Equation (\ref{diagonal-GEM}) therefore demands that the distribution of sources
is isotropic (i.e. the current vector has the same absolute magnitude along all
three spatial directions); it also restricts the magnitude of $\partial_t^2 \Phi$,
a quantity that does not appear in the other two derived equations. The latter
equation (\ref{off-diagonal-GEM}) dictates that the velocity of the source is
strictly non-relativistic, $|\vec{u}| \ll c$: it is only then that the time 
variation of the vector potential is extremely small, due to the suppression
factor $1/c^2$ on its right-hand-side, and the consistency of the complete
set of derived equations is guaranteed. 

We will finally address the question of the equation of motion of a test particle
propagating in the background (\ref{line-element-GEM}). This has also been derived
and discussed in the literature before \cite{Mashhoon} but only in a very simplified
form. Here, we keep all corrections and comment on their importance at the
final stage of the calculation.  As demonstrated above, the consistency of the
set of derived equations dictates that we work in the non-relativistic
limit, in which case we may write
\beq
ds^2 = c^2dt^2-(dx^1)^2 -(dx^2)^2 -(dx^3)^2=
c^2 dt^2 \left(1-\frac{|\vec{u}|^2}{c^2}\right) \simeq c^2 dt^2\,.
\eeq
Then, the spatial components of the geodesics equation
\beq
\frac{d^2 x^\rho}{ds^2} + \Gamma^\rho_{\mu\nu}\,\frac{dx^\mu}{ds}
\frac{dx^\nu}{ds}=0 \label{geodesics-GEM}
\eeq
take the explicit form
\begin{equation}
\frac{d^2x^i}{dt^2}+ c^2\,\G^i_{00}+
2c \,\G^i_{0j}\,\frac{dx^j}{dt}+
\G^i_{kj}\,\frac{dx^k}{dt}\frac{dx^j}{dt} =0\,. \label{geodesics1-GEM}
\end{equation}
In the linear approximation, the Christoffel symbols are given by Eq. (\ref{Christoffel}). 
Reading the form of the initial perturbations $h_{\mu\nu}$ from the line-element
(\ref{line-element-GEM}), we find
\bea
\G^i_{00} &=& 
\frac{1}{c^2}\,\partial_i \Phi
+\frac{2}{c^3}\,\partial_t A^i\,, \\[1mm]
\G^i_{0j} &=& 
\frac{1}{c^2}\,F_{ij} - 
\frac{1}{c^3}\,\delta^i_j \,\partial_t \Phi\,, \\[1mm]
\G^i_{kj} &=& 
-\frac{1}{c^2}\left(\delta^i_j \,\partial_k \Phi +
\delta^i_k \,\partial_j \Phi -\delta_{kj} \,\partial_i \Phi \right),
\eea
where, in the second of the above equations, we have used the definition 
$F_{ij} \equiv \partial_i A_j -\partial_j A_i$. Substituting the above into
Eq. (\ref{geodesics1-GEM}) and using vector notation, we find
\begin{equation}
m\,\vec{a}=\vec{F}=m\,\vec{E}\left(1+\frac{|\vec{u}|^2}{c^2}\right)
+\frac{4m}{c}\,\vec{u}\times \vec{B} +
2m\left[\frac{\vec{u}}{c}\,\frac{\partial_t \Phi}{c} -
\frac{\vec{u}}{c}\,\left(\frac{\vec{u}}{c} \cdot \vec{E}\right)\right],
\label{Lorentz-GEM}
\end{equation}
where, in accordance to the field equations and gauge condition, we have set
$\partial_t {\vec A}=0$ and thus ${\vec E}=-\nabla \Phi$.
A simplified form of the above equation, namely the following
\begin{equation}
m\,\vec{a}=\vec{F}=m\,\vec{E}+\frac{2m}{c}\,\vec{u}\times \vec{B}\,,
\label{Lorentz-Mashhoon}
\end{equation}
is the one that has appeared in the literature before \cite{Mashhoon} expressed,
as claimed, to the lowest order in $\vec{u}/c$, $\Phi$ and $\vec{A}$. As is 
evident from Eq. (\ref{Lorentz-GEM}) we agree on the form of the two
terms, proportional to $m\,\vec{E}$ and $\vec{u}\times \vec{B}$, appearing
in the simplified equation -- the apparent disagreement in the numerical
coefficient of the second term is only due to the different definition
(\ref{EB_GEM}) of the GEM field ${\vec{B}}$ in terms of the vector potential.
Also, the two additional terms proportional to
$\vec{E}$ appearing on the right-hand-side of Eq. (\ref{Lorentz-GEM}) may
also be discarded in the non-relativistic limit in which we work. However, there
is one more term, proportional to the combination $\vec{u}\,\partial_t \Phi/c^2$,
whose suppression of magnitude is not evident. If the time variation of the
scalar potential could be roughly associated with the magnitude of the fluid
velocity, then this term might also be discarded. Nevertheless, apart from
the constraint equation (\ref{diagonal-GEM}) -- which until now was ignored in
the context of gravitoelectromagnetism -- that may hint towards that direction,
there is no reason why this term should not be present. In this aspect, our
results agree with the ones presented in \cite{Costa}\cite{Natario} where the
interested reader may find a more extended analysis on the link between the
time-independence of the GEM potentials and the analogy between gravity
and electro-magnetism.


\subsection{An Alternative ansatz for the metric perturbations}

According to the usual assumptions of General Relativity, the scalar potential
$\Phi$ is only associated with the $\tilde h_{00}$ component of the metric
perturbations.  To preserve the analogy with  electromagnetism, the vector
potential should appear linearly in the expression of $\tilde h_{\mu\nu}$, and 
thus can only be accommodated by the $\tilde h_{0i}$ component. Then, that
leaves $\tilde h_{ij}$ to be either zero, as was the case in the previous
subsection, or to be also associated to the scalar potential $\Phi$. 
Although this may at first seem peculiar, in the course of our analysis it will
be justified and shown to exhibit interesting features. 

To this end, we will now investigate the following alternative assumption
for the perturbations $\tilde h_{\mu\nu}$ 
\beq 
\tilde h_{00}=\frac{\Phi}{c^2}\,, \qquad \tilde h_{0i}=-\frac{A^i}{c^2}\,,
\qquad \tilde h_{ij}=\frac{\Phi}{c^2}\,\delta_{ij}\,. \label{case2}
\eeq
Then, the spacetime line-element assumes the simplified form 
\beq
ds^2=c^2\left(1+\frac{2\Phi}{c^2}\right) dt^2-\frac{2}{c}\,(\vec{A} \cdot d\vec{x})\,dt
- \delta_{ij} dx^i dx^j\,,
\label{line-element-case2}
\eeq
with the potentials $\Phi$ and $\vec{A}$ having again the same interpretation.

Working as before, we first derive the components of $\tilde h_{\mu\nu}$ in mixed
form  and then, from the field equations (\ref{field_eqs}), we obtain the following
system
\bea
0 = 2 k \rho\,u_0\,u_0\,, &&\label{PoissonCase2}\\[2mm]
\frac{1}{c^2}\,\partial_i\left(\partial_k A^k\right)
-\frac{1}{c^2}\,\delta^{kl}\,\partial_k \partial_l A^i = 2k \rho\,u_0\,u_i\,, &&
\label{4Maxwellcase2}\\[2mm]
-\frac{1}{c^3}\,\partial_t \left(\partial_i A^j +\partial_j A^i\right)
-\frac{2}{c^2}\,\partial_i\partial_j \Phi  \hspace*{5cm} &&\nonumber \\
 \hspace{2cm}
+ \delta_{ij}\left[\frac{2}{c^2}\,\delta^{kl}\,\partial_k\partial_l \Phi + 
\frac{2}{c^3}\,\partial_t (\partial_k A^k)\right] = 2k \rho\,u_i\,u_j\,.&&
\label{ExtraCase2}
\eea

Setting again $k \equiv 8\pi G/c^4$, Eq. (\ref{4Maxwellcase2}) takes a form
similar to that of the fourth Maxwell's equation for both a static scalar
and vector potential, $\partial_t \Phi=\partial_t \vec{A}=0$,
\beq
\vec{\nabla}\,(\vec{\nabla} \cdot \vec{A})
-\nabla^2 \vec{A} = 16\pi G \rho \,\frac{\vec{u}}{c}\,.
\label{4Maxwellstatic2}
\eeq
This equation differs from the exact Maxwell equation by a factor of 4 on the
right-hand-side but, as will see, this will be irrelevant. 
Equation (\ref{PoissonCase2}), that in the previous case gave us Poisson's law,
has now reduced to the trivial result $\rho=0$, which when combined with the fact
that $\partial_t \vec{A}=0$, leads to the demand that $\Phi$ satisfies the
equation 
\beq
\nabla^2\,\Phi=0\,.
\label{Poisson2}
\eeq
Then, we conclude that this choice for the gravitational perturbations leads to
a static model of gravity in vacuum. For $\rho=0$, the right-hand-side of 
Eq. (\ref{4Maxwellstatic2}) also vanishes making the numerical factor
irrelevant. Also, in retrospect, our assumption of non-vanishing $\tilde h_{ij}$
seems justified: although this component is indeed significantly suppressed
in the presence of an energy-momentum tensor of the form $T_{\mu\nu}=\rho u_\mu u_\nu$,
in vacuum all components are of the same order. 

Imposing the transverse gauge condition ${\hh^{\m\n}}_{\4,\n}=0$ results into
two constraints, namely
\beq
\frac{1}{c}\,\an[t]\F + \vec{\nabla}\cdot \vec{A}=0\,,
\qquad \quad \frac{1}{c}\,\partial_t \vec{A} + \vec{\nabla} \Phi=0\,,
\label{gauge-case2}
\eeq
from the temporal and spatial components, respectively. The first is the
well-known Lorentz condition, while the latter demands the vanishing of
the GEM field $\vec{E}$. Note that the different numerical coefficients that
appear in the ansatz (\ref{case2}), compared to the traditional one (\ref{case1})
used in GEM, allows us to define the GEM fields $\vec{E}$ and $\vec{B}$  in
an exact analogy with the electromagnetism
\beq
\vec{E} \equiv -\frac{1}{c}\,\partial_t \vec{A} -
\vec{\nabla} \Phi\,, 
\qquad \vec{B} \equiv \vec{\nabla} \times \vec{A}\,.
\eeq

Although this case seems to be not particularly rich in content compared
to the one studied in the previous subsection, it is in contrast free of
additional constraints. The set of equations (\ref{ExtraCase2}) -- both the
diagonal and off-diagonal components -- are trivially satisfied if one uses
the aforementioned constraints (\ref{gauge-case2}) following from the gauge
conditions.

Turning finally to the geodesics equation, by using the expressions for
the initial perturbations $h_{\mu\nu}$ as these are read in the line-element
(\ref{line-element-case2}), we arrive at particularly simple forms
for the Christoffel symbols
\beq
\G^i_{00} = 
\frac{1}{c^2}\,\partial_i \Phi
+\frac{1}{c^3}\,\partial_t A^i\,, \quad
\G^i_{0j}= 
\frac{1}{2c^2}\,F_{ij}\,,  \qquad \G^i_{kj} =0\,.
\eeq
Substituting these into the geodesics equation (\ref{geodesics1-GEM}), we
obtain the exact functional analogue of the Lorentz force
\begin{equation}
m\,\vec{a}=\vec{F}=m\,\vec{E}+\frac{m}{c}\,\vec{u}\times \vec{B}\,.
\label{Lorentz-alter}
\end{equation}
Note that the same coefficient appears in front of the `electric' and
`magnetic' terms, in exact analogy to electromagnetism.
We also observe than no additional terms, not even sub-dominant ones in the
non-relativistic limit, emerge in this case. We therefore conclude that
this particular ansatz for the gravitational perturbations may lead to
another class of phenomena observed in vacuum in the context of
gravitoelectromagnetism: the field equations in conjunction to the gauge
condition lead to a self-consistent set of fundamental equations with
no additional constraints and a remarkable similarity to the 
corresponding formulae of electromagnetism.


\section{True Electromagnetism?}

In the context of gravitoelectromagnetism, the scalar $\Phi$ and vector
potential $\vec{A}$, as well as the corresponding GEM fields $\vec{E}$ and $\vec{B}$,
satisfy equations that are remarkably similar to the ones of electromagnetism.
The question then follows of how much an accurate theory of true
electromagnetism one could obtain starting from a tensorial theory similar
to that of General Relativity and introducing the scalar $\hat{\Phi}(x^\mu)$ and
vector $\hat{\vec{A}}(x^\mu)$ electro-magnetic potentials again via the metric
perturbations $h_{\mu\nu}$. As already mentioned in the Introduction,
the objective of such an analysis would of course
be not to replace Maxwell's theory of electromagnetism by a tensorial theory
but rather to investigate how far the analogy between the two dynamics can go. 

From the analysis of the two subsections of Section 3, it is evident that
the form of the metric perturbations $h_{\mu\nu}$ does affect the equations
for the scalar and vector potentials that are derived from the linearised
Einstein's equations. In order to collectively study a large class of 
choices, we will use the following general form of $\tilde h_{\mu\nu}$
\beq 
\tilde h_{00}=\frac{\alpha \hat\Phi}{c^2}\,, \qquad 
\tilde h_{0i}=-\frac{\beta \hat A^i}{c^2}\,,
\qquad \tilde h_{ij}=\frac{\gamma \hat \Phi}{c^2}\,\delta_{ij}\,, \label{h-EM}
\eeq
where $(\alpha,\beta,\gamma)$ are arbitrary numerical coefficients
\footnote{Contrary to what happens in the context of GEM, where
the numerical coefficient in the expression of $\tilde h_{00}$ is fixed so that
$h_{00}=2\Phi/c^2$ in accordance with the Newtonian limit, here, the coefficient
$\alpha$ can be arbitrary.}. The two
ansatze used in subsections 3.1 and 3.2 correspond to the choices
$(\alpha=\beta=1,\gamma=0)$ and $(\alpha=\beta=\gamma=1)$, respectively.
If one hopes to reduce the gravitational field equations to Maxwell's
equations, the chosen $\tilde h_{\mu\nu}$ must be linear in the electromagnetic
potentials and should not contain any derivatives. The above leads to the
conclusion that the scalar $\hat \Phi$ and the vector $\hat{\vec{A}}$ potential
can be accommodated in the scalar $\tilde{h}_{00}$, vector-like $\tilde{h}_{0i}$
and tensor-like $\tilde{h}_{ij}$ components in a limited number of distinct ways
-- in fact, all the allowed choices are included in our ansatz (\ref{h-EM}). 
The above general form will allow us to carry out a generalised analysis and
investigate the contribution that each component of $\tilde h_{\mu\nu}$ 
would have to the field equations. 

Starting again from the linearised form of Einstein's field equations 
(\ref{field_eqs}), but allowing now for a general coefficient $k$ different
from the usual one $8\pi G/c^4$, and employing the general ansatz (\ref{h-EM}),
we arrive at the following set of equations
\bea
\frac{(\alpha-\gamma)}{c^2}\,\delta^{ij}\,\partial_i \partial_j \hat \Phi =
2 k \rho\,u_0\,u_0\,, &&\label{PoissonCase4}\\[1mm]
\frac{(\alpha-\gamma)}{c^3}\,\partial_i \partial_t \hat \Phi +
\frac{\beta}{c^2}\,\partial_i\left(\partial_k \hat A^k\right)
-\frac{\beta}{c^2}\,\delta^{kl}\,\partial_k \partial_l \hat A^i = 2k \rho\,u_0\,u_i\,, &&
\label{4Maxwellcase4}\\[1mm]
\hspace*{-2cm}-\frac{\beta}{c^3}\,\partial_t \left(\partial_i \hat A^j +
\partial_j \hat A^i\right)
-\frac{2\gamma}{c^2}\,\partial_i\partial_j \hat \Phi \nonumber \hspace*{5cm} &&\\
\hspace*{0cm}
+ \,\delta_{ij}\left[\frac{2\gamma}{c^2}\,\delta^{kl}\,\partial_k\partial_l \hat\Phi + 
\frac{2\beta}{c^3}\,\partial_t (\partial_k \hat A^k)+
\frac{(\alpha-\gamma)}{c^4}\,\partial^2_t \hat \Phi\right]
= 2k \rho\,u_i\,u_j\,.&&
\label{ExtraCase4}
\eea

Equation (\ref{PoissonCase4}) reveals that, for $\alpha=\gamma$, we inevitably
obtain a model of electro-magnetism in vacuum, i.e. with $\rho=0$, as it was 
also found in subsection 3.2 in the context of GEM. The reason for that is that,
for $\alpha=\gamma$, the $\tilde h_{00}$ and $\tilde h_{ij}$ components, that
are both associated with $\hat \Phi$, have contributions to the $(00)$-component
of the perturbed field equations that are of equal magnitude but of opposite sign. 
For $\alpha \neq \gamma$, on the other hand, we obtain Poisson's law 
\beq
\nabla^2\,\hat\Phi=-4\pi \rho\,,
\label{Poisson}
\eeq
under the identification
\beq
k \equiv - \frac{2\pi}{c^4}\,(\alpha-\gamma)\,. \label{k-EM}
\eeq
Adopting the above value for $k$, and using vector notation,
Eq. (\ref{4Maxwellcase4}) in turn can be written in the form 
\beq
\vec{\nabla}\,(\vec{\nabla} \cdot \hat{\vec{A}} + 
\frac{1}{c}\,\partial_t \hat \Phi)
-\nabla^2 \hat{\vec{A}} = 4\pi \rho \,\frac{\vec{u}}{c}\,,
\label{4Maxwellstatic}
\eeq
under the assumption that $\beta=\alpha-\gamma$. Therefore, there are
apparently an infinite number of choices one could make concerning the
numerical coefficients appearing in the perturbations $\tilde h_{\mu\nu}$
and still be able to derive the electromagnetic equations from the field
equations (\ref{field_eqs}). Recalling the usual definitions of the electric
$\hat{\vec{E}}$ and magnetic field $\hat{\vec{B}}$ in terms of the
potentials, i.e.
\beq
\hat{\vec{E}} \equiv -\frac{1}{c}\,\partial_t \hat{\vec{A}} -
\vec{\nabla} \hat\Phi\,, 
\qquad \hat{\vec{B}} \equiv \vec{\nabla} \times \hat{\vec{A}}\,, \label{EB_defs}
\eeq
one may easily see that Eqs. (\ref{Poisson}) and (\ref{4Maxwellstatic}) are
the first and fourth, respectively, Maxwell's equations, namely
\beq
\vec{\nabla} \cdot \hat{\vec{E}} = 4\pi \rho\,, \qquad 
\vec{\nabla} \times \hat{\vec{B}}=\frac{1}{c}\,\partial_t \hat{\vec{E}} + 
\frac{4\pi}{c}\,\vec{j}\,,
\eeq
under the constraint that the vector potential is again static, i.e.
$\partial_t \hat{\vec{A}}=0$, and for $\vec{j}=\rho \vec{u}$. Note, that,
by using the definitions (\ref{EB_defs}) for the electric and magnetic
fields, the remaining two Maxwell's equations follow automatically - in this 
case, these have the form:
\beq
\vec{\nabla} \times \hat{\vec{E}}=-\frac{1}{c}\,\partial_t \hat{\vec{B}}=0\,,
\qquad \vec{\nabla} \cdot \hat{\vec{B}} = 0\,. \label{otherMaxwells}
\eeq

We would also like to note that due to the numerical factor $(\alpha-\gamma)$
appearing in Eqs. (\ref{PoissonCase4}) and (\ref{4Maxwellcase4}),
Maxwell's equations remain unchanged under the simultaneous changes
$(\alpha \leftrightarrow \gamma)$ and $\hat \Phi \rightarrow -\hat \Phi$. 
Therefore, contrary to what happens in GEM where $\tilde h_{00}$ is necessarily
tied to the Newtonian potential and thus should always be non-vanishing,
here, the whole set of Maxwell's equations could also be recovered e.g.
in the simple case where $\tilde h_{00}$ is zero and the potential $\hat \Phi$
is introduced solely through the $\tilde h_{ij}$ component. 
However, as in GEM, Maxwell's equations are accompanied by an additional
set of equations, Eqs. (\ref{ExtraCase4}), where this symmetry is broken: for the
chosen values of $k$ and $\beta$, these take the form
\bea
\hspace*{-2cm}\frac{1}{c}\,\partial_t \left(\partial_i \hat A^j +\partial_j \hat A^i\right)
+\frac{2\gamma}{\alpha-\gamma}\,\partial_i\partial_j \hat \Phi \nonumber \hspace*{5cm} &&\\
\hspace*{1cm}
- \,\delta_{ij}\left[\frac{2\gamma}{\alpha-\gamma}\,\delta^{kl}\,\partial_k\partial_l \hat \Phi + 
\frac{2}{c}\,\partial_t (\partial_k \hat A^k)+
\frac{1}{c^2}\,\partial^2_t \hat \Phi\right]
= 4\pi \rho\,u_i\,u_j/c^2\,.&& \label{additional4}
\eea
For $i=j$, the above reduces to the relation
\beq
\partial_t^2 \hat \Phi =- \frac{4\pi}{3} \rho\,|\vec{u}|^2\,,
\label{diagonalcase1}
\eeq
with $(u^1)^2=(u^2)^2=(u^3)^2$, for all values of $\alpha$ and $\gamma$.
For $i \neq j$, the term proportional to $\delta_{ij}$ in Eq. (\ref{additional4})
vanishes - the remaining terms on the left-hand-side
of the equation can be combined to form the components of the electric field
under the assumption that $\alpha=2\gamma$. In that case, the constraint reads
\beq
\partial_i \hat E_j+\partial_j \hat E_i=- 4\pi \rho\,u_i\,u_j/c^2\,.
\label{off-diagonalcase4}
\eeq
In the opposite case, $\alpha \neq 2\gamma$, the constraint is imposed independently
on the time and space-derivatives of the electromagnetic potentials $\hat{\vec{A}}$ and
$\hat \Phi$. In both cases, the right-hand-side is suppressed by  a $u_i u_j/c^2$ factor,
that, in the no-relativistic limit, is always small.

Therefore, although the four Maxwell's equations are correctly recovered,
these are supplemented by additional constraints resulting from the remaining
components of the linearised field equations - the situation is thus similar
to the one encountered in section 3.1. The additional constraints dictate that
the distribution of charges should be again isotropic. Also, here, the form
of the derived equations match the ones of Maxwell's equations under the 
assumption that the vector potential $\vec{A}$ be static.

Imposing a gauge condition yields additional constraints to the model; some
of them act complimentary completing the theory, however, others lead to
unnecessary restrictions to the electromagnetic potentials. If we consider
again the transverse gauge condition $\tilde h^{\mu\nu}_{\4,\n}=0$, then its
time-component takes the explicit form
\begin{equation}
\frac{\alpha}{c}\,\partial_t \hat \Phi+ \beta\,\partial_i \hat A^i=0\,.
\end{equation}
For all cases with $\alpha=\beta$, the above reduces exactly to the
Lorentz condition 
\begin{equation}\label{Lor_cond}
\frac{1}{c}\,\partial_t\hat \Phi + \vec{\nabla}\cdot \hat{\vec{A}}=0\,.
\end{equation}
Its spatial components, though, lead to the unusual constraint
\begin{equation}\label{gauge_add}
\frac{\beta}{c}\,\partial_t \hat{\vec{A}} + (\alpha-\beta)\,\vec{\nabla} \hat \Phi=0\,,
\end{equation}
where the relation $\beta=\alpha-\gamma$ has again been used. Here, the choice
$\alpha=\beta$ leads to the time-independence of the vector potential $\hat{\vec{A}}$,
a demand that was already evident from the field equations. For the choice
$\alpha=2\beta$, however, this condition demands that the electric field itself 
$\hat{\vec{E}}=-\partial_t \hat{\vec{A}}/c -\vec{\nabla} \hat \Phi$
should vanish. For all other cases, the above equation
imposes an additional constraint between the scalar and vector potentials
which is not present in the traditional electromagnetism.  
The choice of the gauge condition is of course not unique, therefore the
above analysis is by no means exhaustive; it merely acts in an indicative
way regarding the type of the constraints that one would end up with.
Another usual gauge condition, the {\it transverse-traceless} one is further
supplemented by the demand that the trace of $\tilde{h}$ should be zero
which eliminates altogether the scalar potential $\hat \Phi$ from the theory.

We now turn to the equation of motion that a test particle would satisfy in
the context of the same formalism. We will conjecture that a modified
geodesics equation, of the form
\beq
m\,\frac{d^2 x^\rho}{ds^2} + q\,\Gamma^\rho_{\mu\nu}\,\frac{dx^\mu}{ds}
\frac{dx^\nu}{ds}=0\,, \label{geodesics-EM}
\eeq
can indeed describe the motion of a massive, charged particle inside
the resulting electro-magnetic field. Working again in the linear
approximation, the relevant components of the Christoffel symbols that
appear in Eq. (\ref{geodesics-EM}) take the form
\bea
\G^i_{00} &=& 
\frac{(\alpha+3\gamma)}{4c^2}\,\partial_i \hat \Phi
+\frac{\beta}{c^3}\,\partial_t \hat A^i\,, \\[1mm]
\G^i_{0j} &=& 
\frac{\beta}{2c^2}\,\hat F_{ij} - 
\frac{(\alpha-\gamma)}{4c^3}\,\delta^i_j \,\partial_t \hat \Phi\,, \\[1mm]
\G^i_{kj} &=& 
-\frac{(\alpha-\gamma)}{4c^2}\left(\delta^i_j \,\partial_k \hat \Phi +
\delta^i_k \,\partial_j \hat \Phi -\delta_{kj} \,\partial_i \hat \Phi \right),
\eea
where, we have defined the electro-magnetic field-strength tensor
$\hat F_{ij}=\partial_i \hat A_j -\partial_j \hat A_i$.
Substituting the above into Eq. (\ref{geodesics-EM}), we find
\begin{eqnarray}\label{Lorentz-gen}
&& \hspace*{-0.5cm}
m\,a^i+q \left[\frac{\beta}{c}\,\partial_t \hat A^i + \frac{(\alpha+3 \gamma)}{4}\,
\partial_i \hat\Phi\right] - \frac{\beta q}{c}\,\epsilon_{ijk}\,u^j \hat B_k \nonumber \\[1mm]
&& \hspace*{0.5cm} 
- \frac{(\alpha-\gamma)q}{2 c^2}\,(\partial_t \hat \Phi)\,u^i +
\frac{(\alpha-\gamma) q}{4c^2}\left[\partial_i \hat \Phi\,(u^k u^k) -
2(\partial_j \hat \Phi\,u^j)\,u^i\right]=0\,.
\end{eqnarray}
Under the assumption that a robust theory of electromagnetism may follow by
applying the formalism of General Relativity, the above formula stands for
the generalised equation of motion for a massive, charged test particle,
that at the same time defines the Lorentz force of the system.
We note that in the special case where $\alpha=\gamma$, where we
recover a static electromagnetic theory in vacuum~\footnote{For $\alpha=\gamma$,
all Maxwell's equations in vacuum are correctly reproduced from Eqs. (\ref{PoissonCase4})
and (\ref{4Maxwellcase4}) under the identification $k \equiv -2\pi \beta/c^4$ 
and for arbitrary $\alpha$ and $\gamma$.}, all terms in the second
line of the previous equation trivially vanish; then, we readily obtain
the minimal form 
\begin{equation}
m\,\vec{a}=\vec{F}=q\alpha\,\hat{\vec{E}}+\frac{q\beta}{c}\,\vec{u}\times \hat{\vec{B}}\,.
\label{Lorentz}
\end{equation}
In this case the constraint $\beta=\alpha-\gamma$ does not hold, thus we are
free to choose a perturbation configuration with $\alpha=\beta$ in order to
restore a common numerical factor in front of the electric and magnetic terms.
For the particular choice $\alpha=\beta=1$, this factor disappears leaving
behind the exact expression of the Lorentz force -- alternatively, it could
be eliminated by appropriately modifying the original form of the geodesics
equation (\ref{geodesics-EM}).

For $\alpha \neq \gamma$, on the other hand, it is evident that in trying
to build a more realistic electromagnetic theory, i.e. a theory not in vacuum,
the expression of the Lorentz force is bound to obtain corrections to its
traditional form. Applying the constraint $\beta=\alpha-\gamma$, that, as
found above, is necessary to restore the form of Maxwell's equations,
the expression of the Lorentz force is written as
\begin{equation}
m\,\vec{a}=\vec{F}=\frac{q}{4}\,\hat{\vec{E}}\left[(4\alpha-3\beta)+
\frac{\beta\,|\vec{u}|^2}{c^2}\right]
+\frac{\beta q}{c}\,\vec{u}\times \hat{\vec{B}} +
\frac{\beta q}{2}\left[\frac{\vec{u}}{c}\,\frac{\partial_t \hat\Phi}{c} -
\frac{\vec{u}}{c}\,\left(\frac{\vec{u}}{c} \cdot \hat{\vec{E}}\right)\right].
\label{Lorentz-case4}
\end{equation}
The form of the additional terms appearing in the expression of the Lorentz
force is similar to that arising in the context of GEM. The second and 
fifth term can be safely ignored in the non-relativistic limit since
their magnitude is suppressed by a factor of $u^2/c^2$, while the role
and magnitude of the fourth term should be more carefully looked at.
The exact numerical coefficients appearing in front of the various terms
differ depending on the particular choice for the ($\alpha$, $\beta$)
coefficients. Particular configurations leading
to a common numerical factor in front of the dominant electric and
magnetic terms can always be found (e.g. $\alpha=7\beta/4$) while a
remaining overall factor could again be eliminated by the modification
of the original geodesics equation.


\section{Discussion and Conclusions}

There is undoubtly a striking similarity between the gravitational and electromagnetic
forces at classical level.  Even in the context of the General Theory of Relativity, 
the dynamics of the gravitational field resembles the one of the electromagnetic,
a result that led to the development of the theory of gravitoelectromagnetism.
The first part of the present analysis focused on the latter theory aiming at
shedding light to particular aspects of it that have not been adequately discussed
in the literature. 

Starting from the perturbed Einstein's field equations at linear approximation and
employing the standard GEM ansatz for the form of the metric perturbations,
we derived the equations that govern the dynamics of the scalar and vector
potentials as well as that of the corresponding GEM fields. We confirmed that 
these equations have the form appearing in the literature but only under the
assumption that the vector potential is time-independent -- otherwise, important
terms are unjustifiably missing and the analogy between gravity and electromagnetism,
that is the corner-stone of GEM, breaks down. In fact, we showed that the
time-independence of the vector potential is dictated by the transverse gauge
condition that is already incorporated in the standard form of the theory. 
The equations for the GEM potentials and fields that resemble the ones of
electromagnetism are, however, supplemented by additional constraints on
their form; these follow from the remaining components of Einstein's field
equations that are usually ignored. Finally, the derivation of the equation
for the Lorentz force revealed the presence of additional terms not all of
which can be ignored in the non-relativistic limit. 

In the context of GEM we proposed an alternative ansatz for the metric
perturbations. The novel feature of this ansatz was the presence of a
non-vanishing $\tilde h_{ij}$ component -- in GEM, this component
is significantly suppressed compared to the other ones, and thus ignored. 
However, this ansatz led to a set of equations similar to the ones for
static electromagnetism in vacuum. Then, our ansatz was justified: in
the absence of an energy-momentum tensor, all components of the
metric perturbations turn out to be of the same magnitude. The ensuing
analysis of this particular ansatz revealed a number of attractive features:
the additional constraints from the field equations trivially vanish while
the geodesics equation takes the exact form of the equation of the Lorentz
force. 

The similarity in the dynamics of the gravitational and electromagnetic forces
created also great expectations that their unification in the context of a
common theory would be straightforward. However, centuries later, such
a theory is still missing. Motivated by GEM, we posed the question whether
a tensorial theory based on the formalism of General Relativity could exactly
re-produce the theory of electromagnetism, and in the second part of this
work we tried to answer this question. As noted in the Introduction, our
analysis does not aim at replacing the theory of electromagnetism but at
investigating how far the analogy between gravity and electromagnetism
extends.

We employed again Einstein's field equations, with the gravitational constant
replaced by an, initially, undetermined one; that constant was later defined
in terms of the velocity of light. The scalar and vector electromagnetic potentials
were introduced via the form of the metric perturbations. We used a general
ansatz and demonstrated that the field equations indeed reduce to the exact
Maxwell's equations for a large class of choices. However, this formulation of
electromagnetism has a number of weak points, similar to those appearing in
the context of GEM, namely: (i) additional constraints emerge from the field
equations, (ii) employing a modified
geodesics equation, the expression of the Lorentz force may be derived but it
contains extra terms that are not always negligible, (iii) the type of gauge
condition that one may choose to impose on the metric perturbations also
yields new constraints on the electromagnetic potentials, and (iv) the vector
potential should always be time-independent. 

Our present analysis will hopefully work towards opening new directions of
thinking in the context of gravitoelectromagnetism. But we also envisage
that it will provide the basis for the formulation of a more sophisticated
mathematical formalism, inspired by gravity, capable of re-producing the
theory of electromagnetism. Such a formalism would link the two theories
at a deeper level and work towards the formulation of a unified theory. 
Further work is clearly necessary: more thought should be devoted on whether
the additional constraints derived from the field equations carry a physical
content or not; the role of the gauge condition and its effect on the ensuing
form of equations should be looked at; finally, the absence of time-dependence
in the vector potential should be cured - preliminary studies show that a 
partial breaking of the symmetry $h_{0i}=h_{i0}$ for the spatial-temporal
component of metric perturbations may solve this problem. 
We hope to return to these questions in a future work.


\begin{acknowledgements}
The authors would like to thank Luis Filipe Costa and Jose Natario for useful
comments on our manuscript and for communicating
to us important information on previous analyses in this topic.
This research has been co-financed by the European
Union (European Social Fund - ESF) and Greek national funds through the
Operational Program ``Education and Lifelong Learning'' of the National
Strategic Reference Framework (NSRF) - Research Funding Program: 
``THALIS. Investing in the society of knowledge through the European
Social Fund''. Part of this work was supported by the COST Actions MP0905
``Black Holes in a Violent Universe'' and MP1210 ``The String Theory Universe''.

\end{acknowledgements}



\end{document}